\documentclass[aps,prd,showkeys,nofootinbib,superscriptaddress]{revtex4-2}

\usepackage{enumerate}
\usepackage{amsmath,slashed,tensor,amssymb,epsfig,amsthm,bm,graphicx}
\usepackage[caption=false]{subfig}
\usepackage{hyperref}

\newcommand{\be}{\begin{equation}}\newcommand{\ee}{\end{equation}}
\newcommand{\bea}{\begin{eqnarray}}\newcommand{\eea}{\end{eqnarray}}
\newcommand{\brr}{\begin{array}}\newcommand{\err}{\end{array}}
\newcommand{\G}{\textbf}

\newcommand{\dr}{\mathrm{d}}

\def\ph{\varphi}
\def\lan{\langle}
\def\lf{\left}

\def\non{\nonumber}\def\pa{\partial}\def\ran{\rangle}

\def\ri{\right}
\def\al{\alpha}\def\ga{\gamma}
\def\de{\delta}

\def\la{\lambda}\def\La{\Lambda}\def\si{\sigma}\def\Si{\Sigma}
\def\om{\omega}

\def\1{{_{1}}}\def\2{{_{2}}}

\def\noHe0{:\;\!\!\;\!\!:H_e(0):\;\!\!\;\!\!:}
\def\noHm0{:\;\!\!\;\!\!:H_\mu(0):\;\!\!\;\!\!:}

\def\lan{\langle}
\def\lf{\left}

\def\non{\nonumber}
\def\pa{\partial}\def\ran{\rangle}

\def\ri{\right}

\def\al{\alpha}\def\ga{\gamma}
\def\de{\delta}

\def\la{\lambda}
\def\La{\Lambda}\def\si{\sigma}\def\Si{\Sigma}
\def\om{\omega}

\def\1{{_{1}}}\def\2{{_{2}}}
\def\I{{_{\rm{I}}}}\def\II{{_{\rm{II}}}}

\begin{document}

\title{Quantum black holes as classical space factories}

\author{A.~Iorio}
\email{alfredo.iorio@mff.cuni.cz}

\affiliation{Institute of Particle and Nuclear Physics, Faculty  of  Mathematics  and  Physics, Charles  University, V  Hole\v{s}ovi\v{c}k\'{a}ch  2, 18000  Praha  8,  Czech  Republic.}

\author{L.~Smaldone}
\email{Luca.Smaldone@fuw.edu.pl}

\affiliation{Institute of Theoretical Physics, Faculty of Physics, University of Warsaw ul. Pasteura 5, 02-093 Warsaw, Poland.}

\begin{abstract}
Space and matter may both be manifestations of a single fundamental quantum dynamics, as it may become evident during black-hole evaporation. Inspired by the fact that quantum electrodynamics underlies the classical theory of elasticity, that in turn has a natural and well-known geometric description in terms of curvature and torsion, related to topological defects, here we move some necessary steps to find the map from such fundamental quantum level to the emergent level of \textit{classical space} and \textit{quantum matter}. We proceed by adapting the \textit{boson transformation method} of standard quantum field theory to the quantum gravity fundamental scenario and successfully obtain the emergence of curvature and torsion, our main focus here. In doing so we have been able to overcome difficult issues of interpretation, related to the Goldstone modes for rotational symmetry. In fact, we have been able to apply the boson transformation method to disclinations, to relate them to the spin structure and to give an heuristic derivation of the matter field equation on curved space. We also improve results of previous work on the emergence of geometric tensors from elasticity theory, as the non-Abelian contributions to the torsion and curvature tensors, postulated in those papers, here emerge naturally. More work is necessary to identify the type of gravity theories one can obtain in this way.
\end{abstract}

\maketitle
\section{Introduction}

The main goal of this paper is to offer general physical arguments and mathematical tools that govern the emergence of \textit{classical} space as a collective manifestation of an underlying, largely unknown, \textit{quantum} dynamics. In doing so, we have in mind black hole (BH) evaporation, where these fundamental quantum entities might realize this picture \cite{Acquaviva:2017xqi,Acquaviva:2020prd}. Nonetheless, the general framework for emergence should remain valid in other settings.

We take inspiration from the classical theory of elasticity, with its gravity-geometry tensors (metric, $g_{ij}$, curvature, $R^i_{jkl}$, torsion, $T^i_{jk}$), whose ultimate dynamical responsible is quantum electrodynamics (QED): a single quantum interaction at the ``fundamental level'' of the classical collective behavior \cite{feynman2011feynman}.

The second part of this program (from classical elasticity to gravity-geometry theory) is actually an old one, as already in the 1980s Kleinert \cite{kleinert1987,Kleinert:1989ky} and in the 1990s Katanaev and Volovich \cite{Katanaev:1992kh} built-up correspondences between 3-dimensional geometry-gravity theories \textit{alla} Cartan, gauge theories of ISO(3), and classical elasticity, especially in the presence of topological defects. Later this went beyond a simple mathematical tool to describe elastic phenomena and became a paradigm for fundamental physics on the emergence of space, known as ``crystal gravity'' \cite{Zaanen:2021zqs}.

Although, of course, we do keep those results in mind and actually improve some of those derivations, our concern is the first part of the program, that is the general mechanism through which the quantum fundamental dynamics produces the classical emergent theory. The second step, from the classical emergent theory to specific types of gravity theories (e.g., General Relativity (GR), Teleparallel Gravity (TG), Conformal/Weyl Gravity (CG)) lies beyond the scope of this work.

According to the quantum gravity (QG) framework we have in mind, both matter and space are emergent phenomena of a fundamental quantum dynamics \cite{Acquaviva:2017xqi} of basic fermionic \cite{Acquaviva:2020prd} constituents that, with Feynman \cite{feynman2011feynman} and Bekenstein \cite{bekenstein2003information}, we call ``$X$ons'' \cite{Iorio_2019} (for a recent review see Ref. \cite{universe8090455}).

The low-energy theory of reference is a quantum field theory (QFT) living in some \textit{target space} but both, the quantum matter and the classical space (i.e. the classical metric, torsion and curvature tensors), are emerging from the same dynamics. The final outcomes for matter and for space are different. Those $X$ons that bosonize, condense and form the lattice making space, whose approximate description is that of the classical continuum we experience at our scales. Those $X$ons that do not bosonize, make what we identify as ``matter '' at our scale\footnote{It is not clear yet how the ``elementary'' bosons of the Standard Model fit in this picture. From a certain point of view they are matter, as they appear to be distinct from space. From another point of view, all of them, but the Higgs, are quanta transmitting the interactions, including gravity whose ``field'' is the space metric. Hence, from the latter point of view, they are more on the side of space than on the side of matter. In this paper, besides this note, we shall not discuss the issue further and matter is always meant to be fermionic.}. There is a big conceptual difference, though, with the standard understanding of matter, because such particles should have properties that result from the interactions of the fundamental $X$ons with the lattice of the bosonized $X$ons, so they are, in fact, \textit{quasiparticles} rather than elementary. That is why we call this QG picture the ``quasiparticle picture'' \cite{Acquaviva:2017xqi, Acquaviva:2017krr}.

In this QG context we shall use a powerful method that allows to obtain a (dynamical, Haag) map \cite{Haag:1955ev, Umezawa:1982nv, Umezawa:1993yq, Greenberg:1994zu,Blasone2011} from the $X$ons theory to the emergent (matter, space)
\be \label{Xtopsig}
X{\rm ons} \, \to \, (\psi, g_{\mu \nu}) \,,
\ee
where, for simplicity, we take $\psi$ to be one kind of fermionic matter and $g_{\mu \nu}$ to be a classical space(time) metric\footnote{For reasons that will be given later here, the emergence of the physical time-coordinate, with the associated metric structure, will not be discussed in this paper.}.

The method is inspired to the quasiparticle picture of Umezawa and collaborators, see, e.g., Refs. \cite{Umezawa:1982nv} and \cite{Umezawa:1993yq} mostly applied there to condensed matter systems, including the mentioned elasticity theory \cite{Wadati1977ASF,Wadati1978FDP,PhysRevB.18.4077}. That method, though, takes as fundamental the QFT from which the higher level structures emerge.

In our case, the fundamental level is discrete and its dynamics is actually unknown. Therefore, we need to cope with a certain amount of approximation, some given by the introduction of an $X$onic \textit{field}, $\Psi (x)$ to describe the set $\Psi_n$, where the discrete label $n$ is traded for the continuum label $x$, and $\Psi$ is fermionic. Once that is assumed, the procedure can be applied and the goals accomplished. 

The key ingredients of Umezawa's approach, that we apply here to a QG context, are
\begin{itemize}
  \item The \textit{dynamical map} (or Haag expansion) that deals with all the emergent quantum excitations, like the quasiparticles.
  \item The \textit{spontaneous symmetry breaking} (SSB), within the underlying QFT, that produces the Nambu--Golstone (NG) bosonic collective excitations, that are phonons for Umezawa and here are quanta of space.
  \item The SSB, in turn, through the \textit{boson transformation method} \cite{Umezawa:1982nv, Umezawa:1993yq}, gives raise to the \textit{classical collective excitations}, necessary to build the theory of elasticity for Umezawa and the ``theory of space'' for us here.
\end{itemize}

Indeed, the boson transformation method explains the formation of macroscopic objects in QFT by the condensation of a large number of bosonic particles in the vacuum. In particular, when the boson fields are gapless, as for NG fields, the method provides solutions of the field equations with topological defects.

\begin{figure}[htbp]
	\centering
		\includegraphics[width=\textwidth]{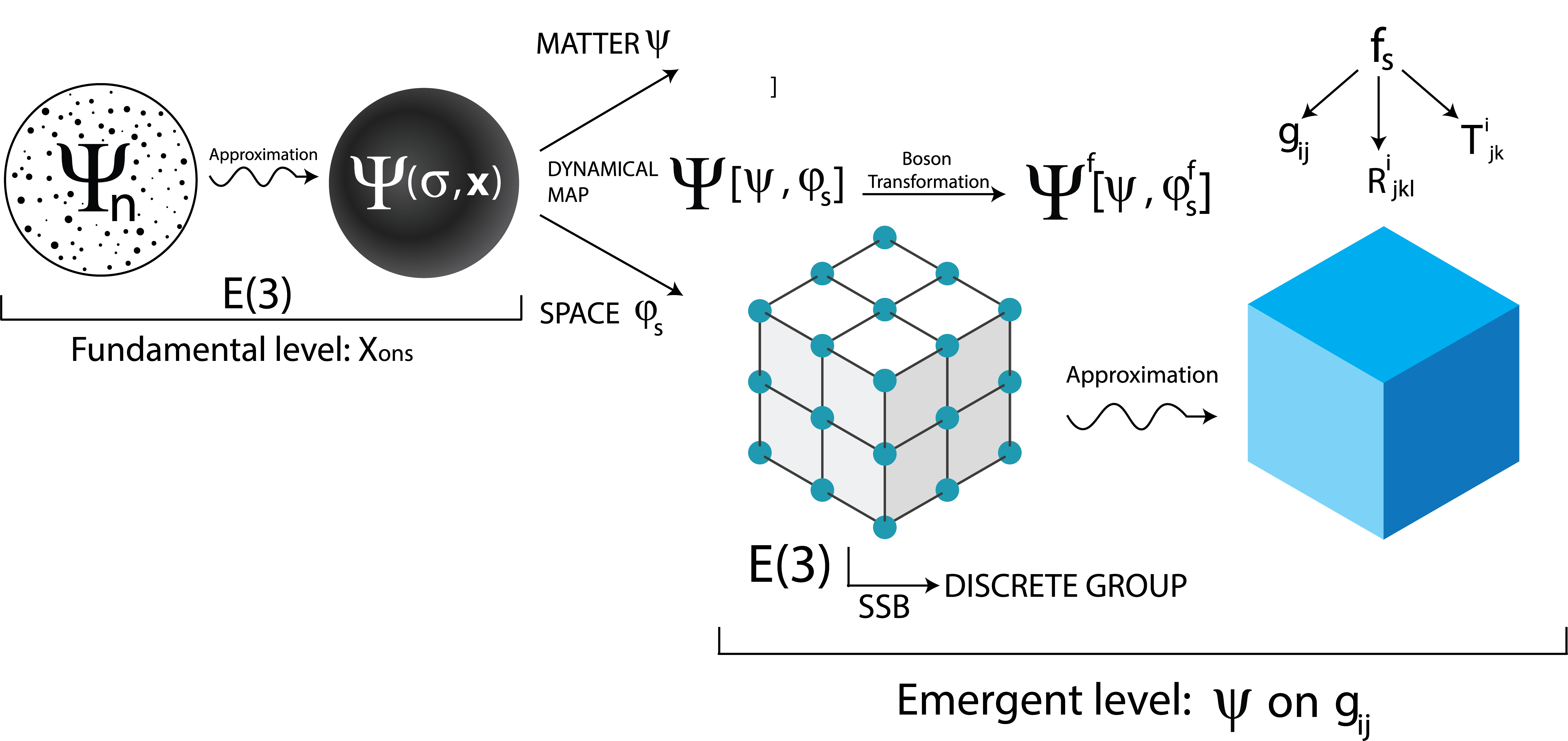}
\caption{A schematic view of the method used in this paper to construct the classical space and its geometric tensor from the underlying quantum gravity. The method is inspired to Umezawa's quasiparticle picture, based on the dynamical (Haag) map and on the boson transformation method in systems with SSB. We have to introduce a couple of approximations, though, the most important being the one that treats the discrete fundamental system as a quantum field $\Psi(x)$.}
	\label{fig:fromxtow}
\end{figure}

In the present work, we implement the boson transformation method to describe the dynamical emergence of a classical geometric structure (metric, curvature, torsion) from an underlying quantum dynamics on a structureless manifold (target space), and NG bosons condensation on vacuum. Let us summarize our approach in a scheme
\vspace{1cm}

\bea \non
{\rm Quasiparticle \ \ picture}
\begin{cases}
 \mathrm{Elasticity \, \,  theory-Classical \, \, geometry} \\[2mm]
\ \ \ \ \ \ \ \ \  \ \ \ \ \ \ \ \ \ \ \Uparrow \ \ ({\rm Crystal \ \ Gravity}) \\[2mm]
\mathrm{Quantum \, \, field \, \, theory \, \, of \, \, crystal-defects} \\[2mm]
\ \ \ \ \ \ \ \ \  \ \ \ \ \ \ \ \ \ \  \Uparrow \ \ ({\rm Boson \ \ method})\\[2mm]
\mathrm{Fundamental \, \, quantum \, \, dynamics} \ \ \ ({\rm QED} \ , \, X{\rm ons})
\end{cases}
\eea
\vspace{1cm}

that, together with the schematic view of Fig.\ref{fig:fromxtow}, should help making clear our method and approximations.

Before starting this journey, let us close this Introduction by putting into context the QG quasiparticle picture we just recalled. The modern approach to QG shifted from the quantization of $g_{\mu \nu}$ as a fundamental field, to the emergence paradigm. According to this, the gravitational field (hence spacetime itself) is a macroscopic or thermodynamics manifestation of the microscopic dynamics. Besides what mentioned, other perspectives are the corpuscular theory of BHs \cite{dvali2011black, Dvali:2012en,Buoninfante:2019fwr} or the SYK model \cite{Nedel:2020bxa}, all strongly inspired by condensed matter physics. Another important line of this type of research is the approach based on the thermodynamics of spacetime, initiated by Jacobson \cite{PhysRevLett.75.1260,PhysRevLett.116.201101,PhysRevD.102.104056,PhysRevLett.96.121301,Padmanabhan_2010,PhysRevD.81.024016}, for which Einstein equations stem from a maximal entropy principle, $\delta S = 0$. Related, but independent, is the approach initiated by Verlinde \cite{Verlinde:2010hp,Visser:2011jp}, that considers gravity as an entropic force.

The paper is organized as follows. In Section \ref{SecXons} we recall the main points of the QG quasiparticle picture of Ref. \cite{Acquaviva:2017xqi} and propose a dictionary to adapt the formalism of the dynamical map to that picture. After a description of the boson method in Section \ref{genfram}, in Section \ref{emspace} we study the SSB of $E(3)$ symmetry, extending some issues discussed in Refs. \cite{Wadati1977ASF,Wadati1978FDP,PhysRevB.18.4077} and applying them to the problem of space geometry and matter emergence from $X$ons dynamics. The last Section is devoted to our Conclusions.

\section{$X$ons, QG quasiparticles and the necessary approximations} \label{SecXons}

The fact that entropy of a local system could be bound \cite{Bekenstein1981} may imply that the fundamental Hilbert space of such local system is finite-dimensional \cite{tHooft:2009fgp,bao2017hilbert,carroll2019mad}. In this case the Hilbert space must describe \textit{all} degrees of freedom (dof), that is those of matter and those of space. BHs are supposed to be the only physical systems which saturate such bounds. The last observation led to the quasiparticle picture of Ref. \cite{Acquaviva:2017xqi}, see also Refs. \cite{Acquaviva:2020prd,Iorio_2019,Acquaviva:2017krr}. In that model, $X$ons indicate the fundamental building blocks whose interaction gives quantum matter and classical space. Their main features are:
\begin{enumerate}
\item Their Hilbert space is finite dimensional.
\item $X$ons forming a BH are free.
\item Their interaction is responsible for space and matter.
\item They are fermions, thus every quantum level can be filled by one mode at most.
\end{enumerate}
In many respects, those features (with the exception, perhaps, of the second) are those of the quantum BH model of Refs. \cite{mukhanov,bekenstein2,bekenstein3}, see also the recent Ref. \cite{progress}. The third point is the one at the basis of BH evaporation phenomenon \cite{Hawking:1974rv,Hawking:1975vcx}: when $X$ons interact they add more space to the space outside the BH (reducing of the horizon area) and produce matter (Hawking radiation) \cite{Acquaviva:2020prd}. Then evaporating BHs are ``factories'' of space and matter. In this paper we shall focus on the first ``product''.

With such picture in mind, in Ref. \cite{Acquaviva:2017xqi} the evolution of the entanglement entropy, between the emergent matter and the emergent space, during BH evaporation, was studied. There it was found that the final BH state has, in general, a nonzero entanglement entropy with the ordinary matter produced. Therefore, seen through the eyes of such ordinary matter, BH evaporation is a nonunitary process, or, in other words, the Page curve does not end at zero entanglement entropy. On the other hand, a unitary evolution is recovered when one analyzes the problem from the more fundamental perspective of the $X$ons, the elementary constituents of both matter and space \cite{Acquaviva:2020prd}. In the latter case, BH evaporation is described by the evolution of the quantum state
\be
|\Psi(\si)\ran \ = \ \prod^N_{i=1} \, \sum_{n_i=0,1} \, (\sin \si)^{n_i} \, (\cos \si)^{1-n_i} \, \lf(a^{\dag}_i\ri)^{n_i} \, \lf(b^{\dag}_i \ri)^{1-n_i}  \, |0\ran_{\I} \otimes |0\ran_{\II} \, ,
\ee
where $a_n$ ($b_n$) are the environment (BH) fermionic ladder operators, acting on the Hilbert space $\mathcal{H}_\I$ ($\mathcal{H}_\II$), so that $a_n|0\ran_\I=b_n|0\ran_\II=0$, $\lf\{a_n, a_m^\dag\ri\}=\lf\{b_n, b_m^\dag\ri\}=\de_{n m}$. The parameter $\si$ is not directly related to the \textit{physical time}, nonetheless it plays the role of an evolution parameter of the BH evaporation, in the sense explained in Ref. \cite{Acquaviva:2020prd}: as $X$ons are supposed to be free in the BH phase, a way to keep track of the BH evolution is a counter that only sees those $X$ons that become interactive, that is precisely the job of $\si$. The maximal entanglement entropy between the modes ``inside'' and the modes ``outside'' the BH, $\mathcal{S}_{max}= N \mathrm{ln} 2$, can be identified with the Bekenstein--Hawking entropy of the BH at $\si=0$, and provides an entropy bound for the BH/radiation thermodynamic entropies.

These considerations are fruitful in the quantum information analysis of BH evaporation, but the crucial question of how space and matter emerge from $X$ons, has not been investigated. Here we would like to take make the most of the powerful machinery of QFT recalled earlier. The problem is that QFT deals with systems with an infinite number of dof while, according to the previous arguments, the dof of a local system like a BH are, in fact, finite. Here comes our first approximation: if we only consider BHs of sizes much bigger that the Planck length, the $X$ons are so many, that a continuum is a good approximation\footnote{Using the methods of Refs. \cite{Acquaviva:2017xqi,Acquaviva:2017krr}, in principle we should be able to estimate the amount of entanglement entropy associated with flat space, hence we should be able to quantify how good is the continuum approximation. These calculations, though, are not easy to perform.}. 

If with $\Psi_n$ we indicate the fermionic dof of the $X$ons, i.e. $a_n$ or $b_n$, the first approximation is then
\be
\Psi_n \rightsquigarrow \Psi(x) \,.
\ee
In other words, we deal with a quantum field depending on $x$ that, in fact, is only a label that identifies the given mode. One way to think of this is that, given the finite set of modes, to swap $\Psi_n \to \Psi_m$ corresponds to $x \to x + s$. Such permutation of the $\Psi$s can be seen as a ``translation'' in the argument $x$. In the symmetric ``phase'', given the indistinguishability of the $\Psi$s, this translation is a symmetry.

Let us stress here that this is indeed an approximation, and perhaps a wild one. We do not know what theory the fundamental dof obey, so we need to make some Ansatz and some approximations, guided by knowing where we have to end-up: quantum fields on classical curved spacetimes, both emerging from one single quantum dynamics. 

If we can live with that approximation, we have many advantages. First, we can proceed nearly as in a standard QFT in ``flat space''. We have to recall, though, that this ``flat space'' coordinate, $x$, is only a formal label, counting where, in the set of $X$ons, that particular $X$on is. Second, we can apply the methods of Refs. \cite{Umezawa:1982nv,Umezawa:1993yq} to let true space emerge, along with its geometric tensors, metric, curvature, tensor and with the quantum matter.

We can now take a step further to say that our starting point is a fermion field
\be
\Psi(x) \ \equiv \ \Psi(\si,\G x) \,.
\ee
described by translationally symmetric equations.
Such field is defined on $\mathbb{R} \times \mathbb{R}^3$, where $\si \in \mathbb{R}$ is the affine parameter above described and $\G x \in \mathbb{R}^3$.

 Actually, the emergence of the physical time-coordinate, with the associated metric structure, will not be discussed in this paper, for two reasons. First, we take the view that time is a measure of the evolution of the entropy at the fundamental level. When the entropy reaches a maximum, as it is supposed to be the case for BHs, then time stops. Henceforth, to obtain an emergent physical time requires the introduction of other ingredients in the quasiparticle picture, not available yet. One road could be Verlinde's view of gravity as an entropic force \cite{Verlinde:2010hp}. The second reason is that we shall obtain the metric (Vielbeine) and the (spin) connection, the key ingredients of the ``theory of space'', by SSB of the appropriate symmetry. It is well known \cite{PhysRevLett.109.160401,PhysRevLett.114.251603} that the SSB of time translation (through which we should be getting the time components) is highly problematic. This as well indicates that for time a road different from that for space is necessary.

As for the other labels, $\G x$, the space $\mathbb{R}^3$ is just an idealized space for those labels that, for \textit{a posteriori} convenience, is taken to be three-dimensional and real, but by no means it is our \textit{physical space} with its classical geometric description! The latter is what we want to see emerging from the underlying quantum theory of $\Psi (\si,\G x)$. It is true that $\mathbb{R}^3$ has a metric structure, but that only means for us that we do have differential operators. We shall argue later that such flat (empty) spaces are, in fact, impossible (unphysical). The metric needs be always $e^a_i \neq \delta^a_i$.

As announced, within the general framework of the SSB of translations and rotations, in what follows we shall use the boson transformation method of Umezawa and collaborators \cite{Umezawa:1982nv, Umezawa:1993yq}. Such method explains the formation of macroscopic objects in QFT via the condensation in the vacuum of a large number of bosons. In particular, when the boson fields are gapless, as for NG modes of the SSB of some symmetry, the method permits to find solutions of the field equations with topological defects. Let us summarize the relevant features of this method, presenting it in a form suitable for our scopes.

Before doing so, it must be stressed, once more, that BH evaporation is the main motivation behind the present work. We tey to shed some light on the mechanism of matter and space formation, which was left to be explained in Refs. \cite{Acquaviva:2017xqi} and \cite{Acquaviva:2020prd}. However, the following considerations are general and could be relevant by themselves, with applications to other contexts, e.g. in condensed matter physics \cite{universe8090455}.
\section{The dynamical map and the boson transformation method} \label{genfram}

\subsection{The dynamical map}
Let us consider an interacting quantum field $\Psi$, satisfying the equation
\be \label{ineqp}
\La(\pa) \, \Psi(x) \ = \ j[\Psi](x) \, ,
\ee
where $\La(\pa)$ is a differential operator, while $j$ is a functional of $\Psi$, describing the interaction. Eq.\eqref{ineqp} can be formally rewritten as\footnote{For notational simplicity, in the following we shall never explicitly write the field renormalization constants.}
\be \label{yfeq}
\Psi(x) \ = \  \Psi_0(x) \ + \ \lf(\La^{-1} \star j[\Psi]\ri)(x) \, ,  \qquad \La(\pa) \Psi_0(x) \ = \ 0 \, ,
\ee
where $\La^{-1} \star j$ is the convolution of the interacting terms with a Green function of the operator $\La$. Notice that such Green's function should be fixed by appropriate boundary conditions. Moreover Eq.\eqref{yfeq} has to be regarded as a weak mapping, i.e. is not true at operator level, but is a relation among matrix elements, on some Hilbert space \cite{greiner2013field}. The Hilbert space where such relation is defined cannot be established a priori and it depends on the physical situation one is going to describe \cite{Blasone:2001aj}. $\Psi_0$ is sometimes called the \emph{physical field}.

In relativistic QFT physical fields are the asymptotic $in/out$ fields and the equation \eqref{yfeq} is known as Yang--Feldman equation \cite{Yang:1950vi}. In condensed matter physics, the name ``quasiparticle fields''is used.  Eq.\eqref{yfeq} can be iteratively solved in terms of a series of normal ordered products of physical fields. Such an expansion is known as \emph{dynamical map} \cite{Umezawa:1982nv, Umezawa:1993yq} or \emph{Haag expansion} \cite{Haag:1955ev,Greenberg:1994zu}.

In general the physical fields content is not known a priori, and its an input of the theory which permits to fix the Hilbert space where the dynamical map is defined \cite{barton1963introduction}. For example, in the case of superconductors the physical excitations above the critical temperature $T_c$ are normal-electrons, while below $T_c$ are the Bogoliubov quasiparticles \cite{PhysRevLett.112.070604}, which describe excitations of the Bardeen--Cooper--Schrieffer (BCS) vacuum. Notice that the BCS and the normal-state vacuum belong to different Hilbert spaces and their respective sets of creation/annihilation operators form unitarily inequivalent representations of canonical anticommutation relations \cite{Miransky:1994vk,Blasone2011}.

An interesting situation is the one considered in Ref. \cite{Blasone:1999aq} in the study of vortex solutions of $\phi^4$ theory. In that case, the asymptotic field $\Psi_0$ is decomposed as
\be \label{bjvortex}
\Psi_0(x) \ = \ \rho(x) \, e^{i \theta(x)} \, ,
\ee
in terms of a massive excitation $\rho$ (radial mode) and the massless NG field $\theta$, that follow coupled equations
\be \label{bjvortexeq}
\La_\rho(\pa) \, \rho(x) \ = \ j_\rho[\rho,\theta](x) \, , \qquad \La_\theta(\pa) \, \theta(x) \ = \ j_\theta[\rho,\theta](x)  \, ,
\ee
equivalent to $\La \Psi_0=0$. From these equations, by means of the boson method described below, the authors derives the classical vortex equations in the Born approximation \cite{kleinert1989gauge}.
In Section \ref{emspace} we shall consider a similar scenario for the emergence of QFT in a curved space. In fact we shall take
\be
\Psi_0(x) \ = \ \Psi_0[x;\psi(x),\ph_1(x), \ldots , \ph_N(x)]  \, ,
\ee
where $\psi$ (the `'radial mode") and $\ph_j$s (NG modes) thus follow coupled field equations
\be \label{bjvortexeq2}
\La_\psi(\pa) \, \psi(x) \ = \ j_\psi[\psi,\ph_1, \ldots , \ph_N](x) \, , \qquad \La_{\ph_j}(\pa) \, \ph_j(x) \ = \ j_{\ph_j}[\psi,\ph_1, \ldots , \ph_N](x)  \, ,
\ee
$j=1,\ldots, N $.
In this paper we have to build matter ($\psi$) and space (stemming from $\ph_j$). Therefore, in the following we shall assume $\psi$ is a massless fermion field, while $\ph_j$s are bosons.
\subsection{The boson transformation method}
The dynamical map can be formally expressed as
\be \label{dmapsi}
\Psi(x) \ = \ \Psi[x;\psi(x),\ph_1(x), \ldots , \ph_N(x)] \, ,
\ee
where $\psi$ and $\ph_j$s are the quasiparticles fields.
On the boson fields we can perform the canonical transformation
\be
\ph_j(x) \ \to \ \ph_j^f(x) \ \equiv \ \ph_j(x) \ + \ f_j(x) \, ,
\ee
with $f_j$ being a $c$-number function. Then, it can be proved that
\be
\Psi^f(x) \ = \ \Psi[x;\psi(x),\ph^f_1(x),\ldots,\ldots,\ph^f_{N}(x)] \, ,
\ee
when $\ph^f_j$ satisfies the same equation as $\ph_j$
\be \label{bjvortexeq3}
 \La_{\ph_j}(\pa) \, \ph^f_j(x) \ = \ j_{\ph_j}[\psi,\ph^f_1, \ldots , \ph^f_N](x)  \, ,
\ee
is still a solution of the original dynamical problem, i.e.
\be
\La(\pa) \Psi^f(x) \ = \ j\lf[\Psi^f\ri](x) \, .
\ee
This is known as \emph{boson transformation theorem} \cite{Matsumoto:1979hs,Umezawa:1982nv}. Note that here we presented a general formulation of the theorem, proved in Refs. \cite{Blasone:1999aq,Blasone:2001aj}, where it is not required $\ph_j$ and $f_j$ satisfy the same (homogeneous) equations. This last (stronger) condition is recovered when $j_{\ph_j}=0$. This general statement of the theorem will permit us to apply the method to the case of coupled equations among $\psi$ and $\ph_j$s.

The solution $\Psi^f$  is appropriate to describe the creation of self-sustained and extended objects, through the condensation in the vacuum of a large number of $\ph_j$ quanta. Relevant example include: topological defects in crystals \cite{Wadati1977ASF,Wadati1978FDP,PhysRevB.18.4077}; solitons \cite{Blasone:2001aj}; vortices in superconductors \cite{LepUme,LepManUme}; strings \cite{TZE197563} and bags \cite{PhysRevD.18.1192} in hadron physics. All those examples are obtained when the $f_j$s are not Fourier-transformable, and this may happen, e.g., when some $f_j$s diverge at $\pm \infty$ or when
\be \label{ctd}
\lf[\pa_\mu \, , \pa_\nu \ri] f_j(x) \ \neq \ 0 \, ,
\ee
at some $x$. This last condition can be only fulfilled when the $\ph_j$s are gapless fields (that is $E(k) \to 0$ when $k \to 0$), as is the case for the NG bosons \cite{Umezawa:1982nv, Umezawa:1993yq}.

Let us assume that the dynamical equations (or the corresponding Lagrangian) for $\Psi$ is invariant under the action of a Lie group $G$, whose generators $t_a$ obey
\be \label{lieal}
\lf[t_a \, , \, t_b \ri] \ = \ i C_{a b c} \, t^c \,,
\ee
with $a,b,c = 1, ..., {\rm ord} G$. After the SSB, some NG fields, $\chi^a(x)$, will appear among the physical fields \cite{Miransky:1994vk}. We shall then consider the boson transformation of NG bosons
\be
\chi^a(x) \ \to \ \chi^a(x) \ + \ f^a(x) \, ,
\ee
that, being gapless, can describe the formation of topological defects \cite{PhysRevD.18.520}.In the next section we shall use such method to generate the space itself as an extended object, while other quasiparticles ($\psi$) will play the role of matter.

This process must not necessarily deal with an infinite number of dof \cite{Acquaviva:2020cjx}, and this is quite important for us. The QFT description only represents an approximation of the real $X$ons with a finite number of dof, as extensively discussed earlier.

A system undergoes SSB when nonzero \emph{order parameters}, i.e. the vacuum expectation values (vevs) of the commutator of the Noether's charges $Q_a$ (quantum generator of the broken quotient group) with some local operators, can be defined. This is also usually expressed  by the less-formal condition
\be
Q_a |0\ran \ \neq \ 0 \, .
\ee
Despite of some technical issues raised by this point \cite{FabPic}, we shall use such condition in the following discussion, because we shall not specify the definition of the order parameters, that is irrelevant for our general considerations.
 The role of $\ph_j$s will be thus acted by the NG bosons.

The non-Abelian formulation of the boson transformation method is based on the observation that, when $\Psi \to \Psi^f$, one has the decomposition \cite{Umezawa:1982nv,Matsumoto:1980uu}
\be \label{psif}
 \Psi^f(x) \ = \  U(x) \, \tilde{\Psi}(x)  \, ,
\ee
where $U(x) \equiv U[f(x)]$ is a local gauge transformation of $G$ and $\tilde{\Psi}$ does not contain $f$ or it only contains $f$ through regular functions (usually through the derivatives $\pa f$, which are assumed to be regular).
To understand why Eq. \eqref{psif} holds, one should remind that when $G$ symmetry is spontaneously broken, the symmetry of the quasiparticle equations is generally described by a different group $G'\neq G$. This phenomenon is known as \emph{dynamical symmetry rearrangement} \cite{Umezawa1965DynamicalRO,Matsumoto:1973hg,Matsumoto:1975fi,DeConcini:1976uk,10.1143/PTP.65.315}. Usually the transformation
\be
\Psi(x) \ \to \ \Psi^c(x) \ = \ e^{i c^a t_a} \, \Psi(x) \, ,
\ee
is dynamically rearranged into the translation symmetry of NG bosons
\be
\chi^a(x) \ \to \ \chi^a(x) \, + \, c^a \, ,
\ee
i.e.
\be
\Psi^c(x) \  = \ \Psi[\chi^a+c^a](x) \ = \ e^{i c^a t_a} \, \Psi(x)
\ee
Because the boson transformation is the `'local" version of such constant boson translation, it induces the local gauge transformation \eqref{psif}, with $U[f]=e^{i f^a(x) t_a}$. It must be stressed that, as we will see below (Eq.\eqref{dynrear}), the case of space translations is peculiar, being also the radial/matter field $\psi$ involved in the dynamical symmetry rearrangement. However, being $\psi$ not involved in the boson transformation, it does not contribute to $U$ and the above argument remains valid in that case. For sake of generality we will write
\be \label{uform}
U(x) \ = \ \exp\lf(-i \, \al^a(x) \, t_a\ri)  \, ,
\ee
where $\al^a(x)$ are functionals of $f^a$.
It is also useful to introduce \cite{Umezawa:1982nv,Matsumoto:1980uu}
\bea
A_\mu(x)       & = & i \, U^{-1} (x) \pa_\mu U (x) \,, \label{am} \\[2mm]
F_{\mu \nu}(x) & = & i \, U^{-1} (x) \lf[\pa_\mu \, , \pa_\nu \ri] U(x) \,. \label{fmn}
\eea
From the above relations it is easy to check that
\be \label{variousf}
F_{\mu \nu}(x) \ = \ \pa_\mu A_{\nu}(x)-\pa_\nu A_{\mu}(x) \, - \, i \, \lf[A_\mu(x) \, , \, A_\nu(x)\ri] \, .
\ee
It is important to notice that it is precisely the singularity of $f^a$y, see Eq. (\ref{ctd}), that allows for nonzero field strength, $F_{\mu \nu}$, even for pure gauge configurations like in (\ref{am}). From Eqs.\eqref{am},\eqref{fmn},\eqref{uform} and \eqref{lieal} we then have

\bea
A_\mu{}^a(x) & = &  \pa_\mu \al^a(x) \, , \\[2mm]
F_{\mu \nu}{}^a(x) & = & \pa_\mu A_{\nu}{}^a(x)-\pa_\nu A_{\mu}{}^a(x) \, +  \, C_{a b c} \, A_\mu{}^{b} (x) \, A_\nu{}^c (x) \, , \non \\[2mm]
& = & \lf[\pa_\mu \, , \, \pa_\nu \ri] \al^a(x) \, +  \, C_{a b c} \,\pa_\mu \al^b(x) \, \pa_\nu \al^c(x) \, , \\[2mm]
A_\mu(x) & = & A_\mu{}^a(x) \, t_a \, , \qquad F_{\mu \nu}(x) \ = \ F_{\mu \nu}{}^{a}(x) \, t_a \, .
\eea
Notice that $F_{\mu \nu}{}^a(x)$ also contains a piece due to the curved structure of the Lie group manifold $G$. The last relation, can be put in a simpler form by introducing the \emph{covariant derivative}
$
D_\mu \ \equiv \ \pa_\mu -i \pa_\mu \al^a(x) \, t_a
$:
\be
F_{\mu \nu}{} \ = \  -i \, \lf[D_\mu \, , D_\nu \ri] \, .
\ee
$A_\mu{}^a$ is a classical emergent gauge field created by the $\chi^a$ condensation in the vacuum. Since $A_\mu{}^a$ and $F_{\mu \nu}{}^a$ have a direct physical interpretation, one requires that
\be \label{bide}
\lf[\pa_\mu \, , \, \pa_\nu\ri] A_\la{}^a \ = \ \lf[\pa_\mu \, , \, \pa_\nu\ri] \pa_\la \al^a (x) \ = \ 0
\ee
that is a Bianchi identity.

The simplest example is the \textit{Abelian} case, that is a $U(1)$-invariant scalar field theory, with SSB. In that case the interacting scalar field $\phi(x)$ is expanded in terms of the asymptotic NG as
\be
\phi(x) \ = \ e^{i \chi(x)} \, F[\pa \chi] \, ,
\ee
where $\chi$ is the NG field (phason).
Performing the boson transformation one gets
\be
\phi^f(x) \ = \ e^{-i f(x)} \, e^{i \chi(x)} \, F[\pa \chi-\pa f] \, .
\ee
Because we assume $[\pa_\mu,\pa_\nu]\pa f=0$. we have
\be
U(x) \ = \ \exp \lf(-i \, f(x)\ri) \, .
\ee
Therefore, $A_\mu=\pa_\mu f$ and $F_{\mu \nu}=[\pa_\mu,\pa_\nu]f$.

For example, a superfluid-like linear vortex/relativistic straight-string along the $z$-axis is formed when $f = n \, \theta$, where $n$ is an integer and $\theta$ is the azimuthal angle in cylindrical coordinates. In that case $A_j = n \pa_j \theta$, $j = 1,2,3$, which is the superfluid persistent current, and $F_{1 2}=-F_{2 1}= 2 \pi n \de(x_1)\de(x_2)$ \cite{Umezawa:1993yq,Acquaviva:2020cjx}.

In the following we shall apply these \emph{general} results to describe the emergence of space and matter. In practice we will study the SSB of $E(3)$ symmetry. In that case, it is precisely the non-Abelian part of the formalism presented above that will play a crucial role to recognize the geometric structures emerging from the non-trivial NG condensation.

\section{Spontaneous breaking of $E(3)$ symmetry and curved space} \label{emspace}

Our goal now is to apply the boson transformation method to the case of the quantum theory of a fermion, $\Psi(x)$, $ x = (\si,\G x) = (\si,x^1,x^2,x^3) \in \mathbb{R} \times \mathbb{R}^3$, that approximates the fundamental $X$ons. We stress once more that $x$ is the continuum limit of a discrete set of indices, while the physical notion of space, equipped with its metric, connection, torsion and curvature, will emerge \textit{after} SSB and NG boson transformation.

We focus on BH evaporation as a phenomenon where space is actually made. Hence, according to Ref. \cite{Acquaviva:2020prd} and to the discussion of Section \ref{SecXons}, we do not have time but rather an affine parameter, $\si$, which counts the interacting $X$ons, i.e., those $X$ons that leave the BH to form space (and matter) ``outside'' the BH. Having no time in this scenario, we cannot define asymptotic \textit{free} fields, as in standard particle physics and our set-up more closely resembles that of condensed matter, where the interaction can never be \emph{turned off}. Moreover, we shall assume that all boson transformations are `'static'', i.e. $f^a(x)=f^a(\G x)$.

Let us take our fermionic system whose Hamiltonian commutes with the six generators of the Lie algebra $e(3)$,
\bea \label{po1}
&& \lf[P_a \, , \, P_b \ri] \ = \ 0 \, ,\\[2mm] \label{po2}
&& \lf[P_a \, , \, J_{b}\ri] \ = \ i \, \varepsilon_{a b c} P_c  \, , \\[2mm] \label{po3}
&& \lf[J_{a} \, , \, J_{b}\ri] \ = \ i \, \varepsilon_{a b c} \, J_{c} \,,
\eea
where $a,b,c=1,2,3$.

Given the dimensions of the problem, the field $\Psi(x)$, that mimics the $X$ons, has two components in the irreducible spinor representation of $SO(3)$
\be
\Psi(x) \ = \ \begin{pmatrix} \Psi_\uparrow(x) \\ \Psi_\downarrow(x) \end{pmatrix} \,.
\ee
Rotations on $\Psi$ then act as
\bea
\Psi'(x) \ = \ e^{i \, \theta^a \, J_a} \, \Psi(x) \, e^{-i \, \theta^a \, J_a} \ = \  e^{-\frac{i}{2} \theta^a \si_a} \, \Psi(\si,\G x') \, , \qquad x'^a \ \equiv \ R^a{}_b(\theta) x^b \, ,
\eea
where $\theta^a$ are real parameters, $R$ belongs to the fundamental representation of $SO(3)$ and $\si_a$ are the Pauli matrices. Moreover, under translations
\be
\Psi'(x) \ = \ e^{i \, u^a \, P_a} \Psi(x) e^{-i \, u^a \, P_a} \ = \ \Psi(\si,\G x') \, , \qquad x'^a \ \equiv \ x^a +u^a
\ee
where $u^a$ are still real parameters.

A crucial point is here the counting of NG modes, that it is known to be not a simple matter, in general \cite{Low:2001bw,Brauner:2010wm,Watanabe:2011ec,Watanabe:2011dk,Watanabe:2013iia,Beekman:2013na,Watanabe:2019xul}.

This is particularly important here, because it is only in the direction where the symmetry is broken that one can produce a nontrivial (classical) gauge field $A_i{}^a$ (that is $e_i{}^a \neq \delta^a_i$ and $\omega_i{}^a \neq 0$) that gives the wanted emergent ``theory of space''.

In Ref. \cite{Watanabe:2013iia} it has been proved that a redundancy of NG modes occurs when
\be  \label{redcon}
\int \!\! \dr^3 x \, \sum_a \, c_a(x) \, j_a^0(x) |0\ran \ = \ 0 \, ,
\ee
where $j_a^0$ are the charge densities corresponding to the broken generators and $c_a$ are generic functions. in our case $P_a=\int \!\! \dr^3 x p_a(x)$, $J_a=\int \!\! \dr^3 x j_a(x)$. $J_a$ can be decomposed as $J_a=L_a+S_a$, with $L_a= \int \!\! \dr^3 x l_a(x)$, $S_a \ = \ \int \!\! \dr^3 x \Si_a(x)$, i.e. the orbital and the spin part, respectively. If we dealt with scalar fields, the spin part was not present and $l_a(x) = \varepsilon_{a b c} x^b p_c(x)$, thus satisfying Eq.\eqref{redcon} in a strong sense. When $S_a|0\ran=0$, Eq.\eqref{redcon} is still fulfilled. In such cases, only NG modes associated to the SSB of translations will be present. However, when $S_a|0\ran \ \neq \ 0$
\be \label{ourcasej}
\int \!\! \dr^3 \, x \, j_a(x)|0\ran \ = \ \int \!\! \dr^3 x \, \varepsilon_{a b c} x^b p_c(x)|0\ran \, + \, \int \!\! \dr^3 x \, \Si_a(x)|0\ran \, .
\ee
Because $\Si^a$ is not linearly dependent on $p^a$, this implies that $j_a$ and $p_a$ are generally linearly independent too. Then, when both $P|0\ran \ \neq \ 0$ and $S|0\ran \ \neq \ 0$, we expect up to six NG bosons.  These considerations will permit us to use the boson method to obtain a non-trivial curvature tensor, and then they represent another argument in favor of the fermionic nature of $X$ons \cite{Acquaviva:2020prd}.

In the next Subsections we shall explore the two possibilities reported in the table. There SSB of translations is always understood\footnote{It is true that one could consider the case $S_a|0\ran \neq 0$, with $P_a|0\ran=0$. However, this is not very interesting because a boson transformation, with the associated formation of an extended object, \emph{always} breaks translation invariance, because the condensate does depend on the position \cite{Umezawa:1982nv,Umezawa:1993yq}.}.
\begin{table}[h]
\label{tabu1}
\caption{Main features of the different SSB phases.}
\begin{tabular}{|c|c|c|}
\hline \hline
 $$ & $ S|0\ran \ = \ 0 $ & $ S|0\ran \ \neq \ 0$\\
\hline
NG fields & $X^a$ & $X^a$ \, , \, $\Theta^a$ \\[2mm]
\hline
 $A_i^a$ &  $ e_i{}^a \ \neq \ \de_i^a \, , \quad \om_i{}^a \ = \ 0 $  & $ e_i{}^a \ \neq \ \de_i^a \, , \quad \om_i{}^a \ \neq \ 0 $   \\
\hline
 $F_{i j}{}^a$ &  $T_{ij}{}^a \ \neq \ 0 \, , \quad R_{il}{}^a \ = \ 0$  & $T_{ij}{}^a \ \neq \ 0 \, , \quad R_{il}{}^a \ \neq \ 0$\\
\hline
\end{tabular}
\end{table}
%
\subsection{Spontaneous breaking of translations} \label{trancase}

We start from the case when $T^3$ translation symmetry is spontaneously broken:
\be
P_a \, |0\ran \ \neq \ 0 \, , \qquad S_a \, |0\ran \ = \ 0
\ee
As previously discussed, we thus expect only three NG modes associated with the fully broken translation symmetry \cite{Wadati1977ASF,Wadati1978FDP,PhysRevB.18.4077,KITAMURA1990509, Watanabe:2011dk,Watanabe:2013iia,Beekman:2013na}). In our case such NG fields are the analogue of the acoustic phonons in a crystal. We indicate them with $X^a(x)$, and we call them \emph{space-phonons}. Moreover, for the sake of simplicity we only consider one type of fermionic matter (quasiparticle) $\psi(x)$.

The breaking could as well be only partial, leaving some directions untouched by the SSB. Say this happens only for one direction, then such direction is a privileged one and the symmetry groups get reduced. This is very much reminiscent of certain approaches to canonical noncommutativity, defined by $[x^i,x^j]= i \theta^{i j}$, where the privileged direction is given by the vector with components $\theta^i = \frac{1}{2} \epsilon^{i j k} \theta_{j k}$. The QFT with reduced symmetry was studied in Ref. \cite{Alvarez-Gaume:2003lup}, see also Refs. \cite{iorio2002space,iorio2008comment}.

As explained in the previous section, the dynamical map is a weak mapping and different Hilbert spaces generally correspond to a different content of physical fields. In other words the same fundamental dynamics can be realized in different ways and boundary conditions are needed to fix the physical representation \cite{barton1963introduction}. In our case we are motivated by the fact that we should recover a QFT in curved space, in some regime. We write the dynamical map as
\be \label{genmap}
\Psi(x) \ = \  \, \Psi_0(x) \, + \, \ldots \, ,
\ee
where $\Psi_0(x)$
\be \label{eqpsi0}
\Psi_0(x) \ = \ \Psi_0[x; \psi(x), X^a(x)] \, .
\ee
decompose in terms of the NG fields $X^a$ (and their derivatives) and of some fermion field $\psi(x)$. Here the dots stands for terms containing higher order normal ordered products of $\Psi_0$. Because $\Psi_0$ satisfies an homogeneous equation $\La(\pa) \Psi_0(x)=0$, $\psi$ and $X^a$ will generally follow some coupled equations (see Eq.\eqref{bjvortexeq2}):
\be \label{coeq}
\La_\psi(\pa) \psi(x) \ = \ j_\psi[\psi,X^a](x) \, , \qquad \La^a_X(\pa) X^a(x) \ = \ j_X[\psi,X^a](x) \, .
\ee
The physical Hilbert space factorizes as $\mathcal{H} \ = \ \mathcal{H}_{X} \otimes \mathcal{H}_{\psi}$.

Let us now perform the boson transformation
\be \label{xbostr}
X^a(x) \ \to \ X_u^a(\G x) \ \equiv \ X^a(x) \, + \, u^a(\G x) \, ,
\ee
with $X_u^a(\G x)$ satisfying the same equation as $X^a(x)$. By means of the Baker--Campbell--Hausdorff formula one can easily verify that such transformation is implemented through an \emph{improper} unitary transformation
\be \label{qtrans1}
X^a_u(x) \ = \ G_X(u) \, X^a(x) \, G_X^\dag(u)  \, , \qquad G_X(u) \ \equiv \ \exp\lf(-i \, \sum_a \,  \int \!\! \dr^3 x  \, u^a(\G x) \, \Pi^a_X(x) \ri) \, .
\ee
where $[X^a(\si,\G x),\Pi^a_X(\si,\G x')=i \de^3(\G x-\G x')]$.
This operation induces an inhomogeneous condensate structure on vacuum
\be
|0(u)\ran \ = \  \, G_X(u) \, |0\ran \, .
\ee
The dynamical map for $\Psi^u$ (the boson transformed $\Psi$ field) then reads
\be
\Psi^u(x) \ = \ G_X(u) \, \Psi(x) \, G_X^\dag(u)  \, \ = \  \Psi^u_0(x) \ + \ \ldots \, ,
\ee
with
\be \label{sumla}
\Psi^u_0(x) \ = \ \Psi_0[x; \psi(x),X^a_u(x)] \, .
\ee

The classical gauge transformation has the general form
\be \label{utransl}
U(\G x) \ = \ \exp\lf(- i \, y^a(\G x) \, \mathcal{P}_a\ri) \, ,
\ee
where $\al^a(\G x)=y^a(\G x)$ is a functional of $u$ so that, in general, $[\pa_i, \pa_j] y^a(\G x) \neq 0$ and
where $\mathcal{P}_a$ are the generators of translations. From now on we shall use the indices $i,j,k$ for $x^i$. $y^a$ has to be regarded as a new set of flat coordinates, while $x^j$ are now curvilinear coordinates. In fact, from Eq.\eqref{am}
\be
A_j(\G x) \ = \ e_j{}^{a}(\G x) \, \mathcal{P}_a \, ,
\ee
with
\be
e_j{}^{a}(\G x) \ \equiv \ \pa_j y^{a}(\G x) \, ,
\ee
being the classical gauge fields emerging from the boson transformation, which can be identified with \emph{Vielbeins} and $\mathcal{P}_a$ form a $2 \times 2$ matrix representation of the $t^3$ Lie algebra of translations in three dimensions. Being these fields physical quantities we should impose the regularity condition  \eqref{bide}
\be
\lf[\pa_i \, , \, \pa_j\ri]e_k{}^a(\G x) \ = \ \lf[\pa_i \, , \, \pa_j\ri]\pa_k y^a(\G x) \ = \ 0 \, .
\ee
Now, as an effect of the condensation, a metric tensor can be defined:
\be \label{gdef}
g_{i j}(\G x) \ \equiv \  \delta_{a b} \, \pa_i y^{a}(\G x) \, \pa_j y^{b}(\G x) \ = \ \delta_{a b} \, e_i{}^{a}(\G x) \, e_j{}^{b}(\G x) \, .
\ee

By means of Eq.\eqref{variousf}, we thus find
\bea
F_{i j}(\G x) \ = \ T_{i j}{}^a(\G x) \, \mathcal{P}_a \, ,
\eea
where
\be
T_{i j}{}^{a}(\G x) \ = \ \pa_i e_j{}^{a}(\G x)-\pa_j e_i{}^{a}(\G x) \, ,
\ee
is the classical torsion tensor. Then we have shown that \emph{classical torsion can emerge from a non-trivial condensate of NG bosons}.

After the boson transformation, one can take vev of the first equation in \eqref{coeq}
with respect to $|0\ran_X$ ($|0\ran \ = \ |0\ran_X \otimes |0\ran_\psi$). This will be an equation which couples $\psi$ with $u$, i.e. the equation for $\psi$ interacting with a classical extended object. Then
\be \label{coequt}
{}_X\lan 0|\La_\psi(\pa) \psi(x)|0\ran_X \ = \ {}_X\lan 0|j_\psi[\psi,X^a_u](x)|0\ran_{X} \, .
\ee
is a QFT equation on a space with torsion.

The detailed form of $y$ clearly depends on the specific model considered. In order to give a relevant example, we shall think about the situation when a crystal structure is formed. In that case $E(3)$ is broken down to the discrete crystalline group:
\be
\psi(x) \ \to \ \psi \left(\si, \G x + \sum m_i \G l_i \right) \,, \qquad \psi(x) \ \to \ \psi(\si, D_R \,  \G x) \, ,
\ee
with $D_R$ being specific discrete rotations and $m_i \in \mathbb{Z}$, $\G l_i$ being the lattice vectors. Because we are thinking about QG we may suppose that $|\G l_i| \approx l_P$, where $l_P$ is the Planck length.
In the study of crystal defects in QFT with the boson transformation method, in Refs. \cite{Wadati1977ASF,Wadati1978FDP,PhysRevB.18.4077} it has been proved that the translation symmetry of the equations for the interacting field (in our case Eq.\eqref{ineqp}), is dynamically rearranged \cite{Umezawa1965DynamicalRO,Matsumoto:1973hg,Matsumoto:1975fi,DeConcini:1976uk,10.1143/PTP.65.315} into the symmetry
\be \label{dynrear}
X^a(x) \ \to \ X^a(\si,\G x+\G c) \, + \, c^a \, , \qquad \psi(x) \ \to \ \psi(\si, \G x+\G c) \, ,
\ee
of the quasiparticle fields equations \eqref{coeq}, with $\G c$ being a constant vector. Then, it can be deduced that the phonon fields appear in the dynamical map of interacting fields in the linear combination $x^a+X^a(x)$ . Then, the boson transformation \eqref{xbostr} induces
\be
 x^a \ \to \ y^a(\G x) \ \equiv \ x^a \, + \, u^a(\G x) \, ,
\ee
the first term being a label, $x^a$, the second term being a field $u^a(\G x)$.
For small $u^a$ one recovers the linearized description of Refs. \cite{kleinert1987,Kleinert:1989ky}
\be \label{gdefl}
g_{i j}(\G x) \ \approx \  \delta_{i j} \, + \, \pa_i u_j(\G x) + \pa_j u_i(\G x) \, ,
\ee
with $u_i(x) \ \equiv \ e_i^{a} u_a(x)$.
Notice that, in such case
the condition \eqref{ctd} indicating the formation of topological defects
\be
\lf[\pa_i,\pa_j\ri] u^a(\G x) \ \neq \ 0 \, .
\ee
is equivalent to $T_{i j}{}^{a}(\G x) \neq 0$.

To compute an Haag map is never an easy task, and in absence of a specific model we can only propose a reasonable ansatz. In the present case:
\be \label{pladynmap}
\Psi_0(x) \ = \ e^{-i \lf(x^a+X^a(x)\ri) \mathcal{P}_a} \, \psi(x) \, .
\ee
In such a way:
\begin{itemize}
\item
$\psi$ is the `'radial mode"
\item
NG enters in the phase of $\Psi_0$, thus appearing from the second order in the complete dynamical map (expanded as functional of $\psi$ and $X^a$). This is expected, being $X^a$ bound states.
\item
$X^a$ appear in the correct combination with $x^a$
\item
The transformation \eqref{dynrear} leads to
\be
 e^{-i \lf(x^a+c^a+X^a(\si, \G x+\G c)\ri) \mathcal{P}_a} \, \psi(\si, \G x+\G c) \ = \ \Psi_0(\si, \G x+\G c) \, ,
\ee
i.e. the transformation \eqref{dynrear} actually corresponds to a translation of a constant vector in $\Psi$, at the linear order in the dynamical map.
\end{itemize}

After the boson transformation (when a non-trivial geometry emerges)
\be \label{psiu0}
\Psi^u_0(x) \ = \ e^{-i \lf(y^a(\G x)+X^a(x)\ri) \mathcal{P}_a} \, \psi(x) \ = \ U(\G x) \, \tilde{\Psi}_0(x) \, ,
\ee
with $\tilde{\Psi}_0 \equiv e^{-i X^a(x) \mathcal{P}_a} \, \psi(x)$. Then, at the linear order in the dynamical map, we managed to confirm that $U(\G x)$ has the form \eqref{utransl}.

If now the differential operator $\La(\pa)$ is linear, $\La(\pa) = \ga^\mu \, \pa_\mu$ ($\mu=0,1,2,3$), with $2 \times 2$ matrix coefficients, one has
\be
{}_X \lan 0|\La (\pa)\Psi^u_0(x)|0\ran_X \ = \ {}_X \lan 0|\ga^a \pa_a \Psi^u_0(x)|0\ran_X \ = \ 0 \, ,
\ee
and then, from Eq. \eqref{psiu0}, employing the normal ordering prescription and that ${}_X\lan 0|:X^a \ldots X^a \pa X^a \ldots \pa X^a:|0\ran=0$, we get
\be
i \, \ga^\mu D_\mu \, \psi(x) \ = \ 0 \, , \qquad D_\mu  \ = \ \lf(\pa_\si , \pa_j+e_j{}^a(\G x) \, \mathcal{P}_a\ri) \, .
\ee
 The above is the dynamical equation of a massless fermion field coupled with torsion in three dimensions. A similar equation governs the dynamics of the conductivity electrons of graphene in the presence of certain topological defects  \cite{IORIO2018265} and enjoys Weyl symmetry, and instance that might have far-reaching consequences \cite{IORIO20111334}.

The classical gauge field $e_j{}^a$ are the quantities we were looking for. They have direct physical consequences and are formed when a large number of NG bosons, $X^a$, condense in vacuum. Then, the vacuum has a coherent-structure, and the classical geometry of space, i.e., the Vielbein $e_j^a$, naturally emerge. When topological defects are present, a nonzero torsion will appear at the classical level. The \emph{bare} vacuum, with zero NG condensed, is an idealization and cannot be regarded as physical. This situation resembles the one of a Bose gas at zero temperature, where the physical vacuum is a Bose--Einstein condensate \cite{Miransky:1994vk}. The case when $u^a(\G x)$ is a regular function corresponds to zero torsion--zero curvature space: Even though the metric could be different from $\de_{ij}$, an observer living on such space cannot distinguish the various situations when $u^a$ are regular. Note that $g_{ij}=\de_{i j}$ when $u^a=constant$. This picture agrees with other scenarios where the flat space vacuum is regarded as a coherent state \cite{Dvali:2015rea}.

The case we just described can only accommodate a teleparallel description of space with nonzero torsion and no curvature, as noticed in Ref. \cite{Yajima:2016gaa} for the Finsler geometry description of defects. In the following we shall show how the formation of a nontrivial spin structure permits to describe the emergence of a non-zero curvature.
\subsection{The emergence of torsion and curvature} \label{comcase}

When
\be \label{comcond}
P_a \, |0\ran \ \neq \ 0 \, , \qquad S_a \, |0\ran \ \neq \ 0 \, ,
\ee
we are spontaneously breaking $E(3)$ and we expect up to six NG fields $X^a(x),\Theta^a(x)$, as explained above. Let us just stress that the presence of rotational NG fields is only possible because of the fact that $X$ons are fermions. A physical way of understanding this fact is to look at condensed matter systems: because the spin structure, \emph{disclinations} (related to curvature) are described in an independent way from \emph{dislocations} (related to torsion), i.e. the angle field $\theta^a(\G x)$ is independent on the deformation parameter $u^a(\G x)$. This is not true otherwise and usually $\theta^a=\varepsilon_{a b c} \pa_b u^c$. In this last case $\theta^a$ represent gaped modes (see e.g. Ref. \cite{Kleinert:1989ky}). This is the physical counterpart of the formal discussion presented at the beginning of this section on the counting of NG modes. Moreover, the following analysis could be viewed as a way of studying disclinations with the boson transformation method, because they were not treated in the original works on this subject (see Refs. \cite{Umezawa1965DynamicalRO,Matsumoto:1973hg,Matsumoto:1975fi,DeConcini:1976uk,10.1143/PTP.65.315}). Notice that the fermionic nature of $X$ons was deduced in a different way in Ref. \cite{Acquaviva:2020prd} where a link between the Pauli principle and the Bekenstain bound on entropy was established.

The dynamical map for the $X$on field should be now of the general form \eqref{genmap}, with
\be \label{eqpsi02}
\Psi_0(x) \ = \ \Psi_0[x; \psi(x), X^a(x), \Theta^a(x)] \, .
\ee
The boson transformation, which describes condensation of NG fields will be now
\be \label{xtbtran}
X^a(x) \ \to \ X^a_u \ \equiv  \ X^a(x) \, + \, u^a(\G x) \, , \qquad \Theta^a(x) \ \to \ \Theta_\theta^a(\G x) \ \equiv \  \Theta^a(x) \ + \ \theta^a(\G x) \, ,
\ee
and
\be
\Psi^{u,\theta}(x) \ = \  \Psi_0^{u,\theta}(x) \ + \ \ldots \, , \qquad
\Psi^{u,\theta}_0(x) \ = \ \Psi_0[x; \psi(x), X^a_u(x), \Theta^a_\theta(x)] \, .
\ee
Transformation \eqref{xtbtran} is formally implemented as in Eq.\eqref{qtrans1} and by
\be \label{qtrans2}
\Theta^a_\theta(x) \ = \ G_\Theta(\theta) \, \Theta^a(x) \, G_\Theta^\dag(\theta) \, , \qquad G_\Theta(\theta) \ \equiv \ \exp\lf(-i \, \sum_a \,  \int \!\! \dr^3 x  \, \theta^a(\G x) \, \Pi_\Theta^a(x) \ri) \, .
\ee
where $[\Theta^a(\si, \G x),\Pi^a_\theta(\si, \G x')=i \de^3(\G x-\G x')]$-
The vacuum describing the condensate is then
\be
|0(u,\theta)\ran \ = \  \, G_X(u) \, G_\Theta(\theta) \, |0\ran \ = \  G_X(u) \, |0\ran_X \otimes G_\theta(\theta) \, |0\ran_\Theta \otimes |0\ran_\psi  \, .
\ee

In this case, the gauge matrix reads
\be \label{utransl1}
U(\G x) \ = \ \exp\lf(-i \, \theta^{a}(\G x) \, \mathcal{J}_a-i \, y^a(\G x) \, \mathcal{P}_a\ri) \, ,
\ee
with $\mathcal{P}_a$ and $\mathcal{J}_a$ form a $2 \times 2$ matrix representation of the $e(3)$ Lie algebra \eqref{po1}-\eqref{po3}.
Then
\be
A_j(\G x) \ = \ \om_j{}^{a}(\G x) \, \mathcal{J}_a \ + \ e_j{}^{a}(\G x) \, \mathcal{P}_a \, .
\ee
Here the classical gauge field $\om_j{}^a$, emerging after the condensation of $\Theta$ modes in the vacuum, can be identified with the \emph{spin connection}.
By means of Eq.\eqref{variousf}, we thus find
\bea
F_{i j}(\G x) \ = \ R_{i j}{}^{a}(\G x) \, \mathcal{J}_a \, + \, T_{i j}{}^a(\G x) \, \mathcal{P}_a \, ,
\eea
where the curvature tensor is given by
\bea
R_{i j}^{a}(\G x) \ = \ \pa_i \om_j{}^{a}(\G x)-\pa_j \om_i{}^{a}(\G x) \ + \ \varepsilon_{a b c} \,  \om_i{}^{b} (\G x) \om_j{}^{c}(\G x) \, ,
\eea
while the torsion tensor acquires a non-Abelian contribution, as it must be for a non zero spin-connection\footnote{Notice, though, that such terms in other approach have to be added by hand.}
\bea
T_{i j}{}^{a}(\G x) \ = \ \pa_i e_j{}^{a}(\G x)-\pa_j e_i{}^{a}(\G x) \ + \ \varepsilon_{a b c} \, e_i{}^{b}(\G x) \, \om_j{}^{c}{}(\G x)  \, ,
\eea
This is the main goal of this paper: \emph{a non-trivial boson condensation of NG fields after the $E(3)$ SSB of a fermion-field theory, will be perceived by a low-energy (continuous) observer as non-zero curvature and torsion of space.}

As in the previous subsection, notice that the condition
\bea \label{onlyR}
R_{i j}{}^a(\G x) & = & \lf[\pa_i , \pa_j\ri] \, \theta^a(\G x) \, + \, \varepsilon_{a b c} \, \pa_i \theta^b(\G x) \, \pa_j \theta^c(\G x) \ \neq \ 0 \, , \\[2mm]
T_{i j}{}^a(\G x) & = & \lf[\pa_i , \pa_j\ri] \, y^a(\G x) \, + \, \varepsilon_{a b c} \, \pa_i y^b(\G x) \, \pa_j \theta^c(\G x) \ \neq \ 0 \, , \label{onlyT}
\eea\
reduces to
\be
\lf[\pa_i , \pa_j\ri] \, \theta^a(\G x) \ \neq 0 \, , \qquad \lf[\pa_i , \pa_j\ri] \, u^a(\G x) \ \neq 0
\ee
for small $u^a$ and $\theta^a$, and in the case of a crystal ( $y^a = x^a+u^a(\G x)$). A World-Crystal like description \cite{kleinert1987,Jizba:2009qf,Zaanen:2021zqs} is thus recovered as a particular case.
Moreover, in such case one can use the same arguments as in the previous subsection, to get
\be
\Psi_0(x) \ = \ e^{-i \lf[\lf(x^a+X^a(x)\ri) \mathcal{P}_a+\Theta_a(x) \mathcal{J}_a\ri]} \, \psi(x) \, ,
\ee
in analogy with Eq. \eqref{pladynmap}.
After the boson transformation
\be
\Psi^{u,\theta}_0(x) \ = \ e^{-i \lf[\lf(y^a(\G x)+X^a(x)\ri) \mathcal{P}_a+\lf(\Theta_a(x)+\theta^a(\G x)\ri) \mathcal{J}_a\ri]} \, \psi(x) \, .
\ee
In the case $\La(\pa) = \ga^\mu \, \pa_\mu$, because ${}_{X,\Theta}\lan 0|:(X^a \mathcal{P}_a+\Theta^a \mathcal{J}_a) \ldots (X^a \mathcal{P}_a+\Theta^a \mathcal{J}_a):| 0\ran_{X,\Theta}=0$ ($| 0\ran_{X,\Theta} \equiv |0\ran_X \otimes |0\ran_\Theta$), one gets
\be \label{psicurved}
i \, \ga^\mu D_\mu \, \psi(x) \ = \ 0 \, , \qquad D_\mu \ = \ \lf(\pa_\si, \pa_j+e_j{}^a(\G x) \, \mathcal{P}_a+\om_j{}^a(\G x) \, \mathcal{J}_a \ri) \, ,
\ee
which is the equation for massless fermions in a three-dimensional curved space with torsion \cite{IORIO2018265}. On the other hand, the spatial part of this equation is what should be written when the matter is coupled to (spatial) three-dimensional gravity
as a gauge theory of the given space(time) group (Poincar\`e, de Sitter, anti de Sitter), as in the classic work of Ref. \cite{WITTEN198846}.

\section{Conclusions}

If space and matter were indeed both manifestations of one fundamental quantum dynamics, then the quantum matter and the classical space where such matter lives, as we experience at our energies, are both emergent phenomena. Finding the map, from the fundamental level to the emergent level, is then crucial to learn how to spot the emergent nature of what we deem to be fundamental.

In this paper we have constructed the building blocks of the ```classical theory of space'': metric, connection, curvature, torsion. We have done that along the lines of what can be done (and was done) for the ``classical theory of elasticity'', where one can define those geometric quantities, especially when topological defects are present. In both cases such classical structures stem from an underlying quantum theory. In the latter case, such quantum theory is QED, that we know quite well. In the former case, such quantum theory is largely unknown and only posited, nonetheless we reached here a series of interesting results.

Here we did that by adapting powerful methods of QFT: the dynamical (or Haag) map that relates fundamental and unreachable quantum fields to measurable quantum fields
\be
\Psi (x) \to \Psi \lf[x;\psi, \ph_1, \dots, \ph_N\ri] \,,\nonumber
\ee
and the boson transformation method, where the bosonic NG quantum fields, $\ph_a$, are shifted by some classical fields $f_a$
\be
\ph_a \to \ph_a + f_a \,.\nonumber
\ee
so that they describe the formation of self-sustained, classical structures, emerging from the underlying quantum dynamics.

The focus here has mainly been on the bosonic part (the fermion $\psi$ being the quantum quasiparticle matter) that arise when the six generators of the $E(3)$ symmetry of the original $X$ons theory are spontaneously broken: three of them giving raise to the wanted Vielbeine
\be
f_a^{transl}=u^a \to e_i{}^a \,, \nonumber
\ee
three of them giving raise to the wanted spin connection $\omega^a_i$, through which we have the building blocks of the emergent ``theory of space'' we are looking for.
\be
f_a^{rot}=\theta^a \to \omega_i{}^a \,. \nonumber
\ee
In order to apply this method we had to clarify the delicate issue of the counting of NG bosons, for the SSB of the Euclidean group $E(3)$, in the presence of a spin structure. In the quantum theory of crystals, it is known
that, despite the generators of $E(3)$ are six, only three NG modes (acoustic phonons) appear in the spectrum, the deep reason residing in Eq. \eqref{redcon}, and introduced in Ref. \cite{Watanabe:2013iia}. Within the boson transformation method, three phonons can only account for dislocation (torsion), but not for disclination (curvature). Here, by making use of the theorem proved in Ref. \cite{Watanabe:2013iia}, we managed to show that, in order to have the necessary three extra modes associated to curvature, the non-trivial spin structure of the vacuum (see Eq.\eqref{ourcasej}) is the key to solve the enigma. Such a problem is a novel one, in the field, because what it is customary done, for instance in Ref. \cite{Kleinert:1989ky}, is to treat disclinations and dislocations as not independent. Henceforth, in that case, three NG modes only are sufficient. That is, clearly, something untenable in a fundamental analysis like the one proposed here where a Cartan-like approach is necessary.

Our starting point is the quasiparticle picture of Ref. \cite{Acquaviva:2017xqi}, that points to fundamental quantum $X$ons of fermionic nature \cite{Acquaviva:2020prd}, and is inspired by analog models of QG constructed on Dirac materials \cite{universe8090455}. Many other scenarios are actually available \cite{gibbons1993euclidean,zwiebach2004first,Rovelli:2004tv,Freidel:2005qe,platania2018asymptotically,Loll:2019rdj}, and some of those give a prominent role to fermions, see e.g. Refs. \cite{HEBECKER2003269,Sachdev:2015efa,Kitaev:2017awl,Nedel:2020bxa,PhysRevA.105.043316,Kaushik:2022bzy}.

In our case, the lattice structure is dynamically generated and not fundamental, as, e.g., in the world crystal model \cite{kleinert1987,Jizba:2009qf,Zaanen:2021zqs}, and we also improve the results of Refs. \cite{Kleinert:1989ky} and \cite{Katanaev:1992kh} in two respects: first,  the non-Abelian part of the torsion and curvature tensors here emerge naturally,
from the boson method for non-Abelian symmetries; second, we can overcome the limits of linearized gravity. Finally, we applied the boson method to disclinations and relate them to the spin structure, something that was not done before.

Remarkably, we also managed to give a derivation of QFT equations for the matter field $\psi$ on the curved space, where it couples with the Vielbein $e_i^{a}$ and the spin connection $\om_{i}^a$ (see Eq. \eqref{psicurved}). In fact, this point presents at least a couple of hard challanges. First, the computation of a dynamical map is not a trivial task, even in a well-defined model, whereas in the present case we do not even have a specific Lagrangian. In order to envisage the general properties of the dynamical map, we had to use arguments based on dynamical symmetry rearrangement \cite{Umezawa1965DynamicalRO,Matsumoto:1973hg,Matsumoto:1975fi,DeConcini:1976uk,10.1143/PTP.65.315}. Second, the quasiparticle fields are usually described as free fields. Then, in order to get the coupled equations of bosons (space) and fermions (matter), we considered the case where $\Psi_0$ follows an homogenoous field equation, while $X^a$, $\Theta^a$ (the phase fields) and $\psi$ (the ``radial mode''), follow coupled equations, as it happens in the description of relativistic vortices \cite{Blasone:1999aq} (see Eqs.\eqref{bjvortex},\eqref{bjvortexeq}). The way out was to employ the \textit{generalized} formulation of the boson transformation theorem (see Refs. \cite{Blasone:2001aj,Blasone2011}), which is quite more complicated than the one customarily used.

There is still a long way to go, though. On the gravity side, we should first include time in this picture. Perhaps, as said, this might happen putting this special dimension next to the ``fundamental'' entropy of entanglement between space and matter, left after the BH evaporation \cite{Acquaviva:2017xqi}. Once this is done, we shall be able to select the kind of gravity theory one can obtain. There are many possibilities, but our world appears to be well described by GR, so we need to keep that in mind.

Then there is the intriguing issue of matter as quasiparticles. Perhaps one direction to investigate with that picture in mind are the oscillations of neutrinos and other particles. Perhaps such ``undecided nature'' of these particles reveals an underlying more fundamental structure? In fact, the QFT description of particle mixing and oscillations requires a condensate structure of vacuum \cite{ALFINITO199591,bigs2,Smaldone:2021mii}, and a dynamical origin of such condensate from an underlying fermionic dynamics has been investigated \cite{Mavromatos:2009rf,Mavromatos:2012us,bigs1}. And how about the bosons of the Standard Model? Are they more on the side of matter or more on the side of `space', thought of as the result of interactions? Finally, $X$ons are natural candidates for dark matter: in their repeated recombination, when BHs are formed and then melt, they can form matter of kinds we have never experienced.

To close, let us say that this work enhances the level of understanding of how to construct classical geometric tensors from the quantum theory, and this may pay back not only in the purely QG scenarios just evoked, but also in applications to condensed matter analogs. In particular, it might help to find the appropriate geometric-gravity dynamical theory of the graphene membrane that is necessary to move forward in the use of that material as an analog of QG \cite{universe8090455,IORIO20111334,Iorio:2011yz,Iorio:2013ifa,IorioReview,IorioPais,IORIO2018265,timeloopPRD2020}.

In fact, the next step in the analog enterprise is to catch BH thermodynamics, and the results obtained here are very helpful in that sense. Indeed, the reproduction of kinematical aspects is like taking a snapshot of two different dynamics at a point of their evolutions when they are the same. This is important and led to many interesting results, but to go beyond that we need that the evolution of the target system and the evolution of the analog system are the same, at least partially. This paper, although fully focused on the QG fundamental scenario, explicitly takes methods and inspirations from the condensed matter world, henceforth the way back, from QG to condensed matter, is actually possible to envisage.

\section*{Acknowledgements}
A.I. acknowledges support from Charles University Research Center (UNCE/SCI/013). L.S. was supported by the Polish National Science Center grant 2018/31/D/ST2/02048 and thanks the Institute of Particle and Nuclear Physics of Charles University for the kind hospitality while parts of this work were completed. The authors thank Giovanni Acquaviva and Pablo Pais for many interesting and informative discussions and Giulio Francesco Aldi for the precious graphical support.

\bibliographystyle{apsrev4-2}
\bibliography{librarySvN}

\end{document}